\documentclass[aps,pre,reprint]{revtex4-1}

\pdfoutput=1

\usepackage{mathtools}	
\usepackage{amssymb}	
\usepackage[italicdiff]{physics}	

\usepackage{xcolor} 		
\usepackage[colorlinks, citecolor={blue!70!black}, urlcolor={blue!70!black}, linkcolor={red!70!black},hyperindex,breaklinks]{hyperref}	

\usepackage{graphicx}	
\usepackage{float}		

\newcommand{\beq}{\begin{equation}}
\newcommand{\eeq}{\end{equation}}

\renewcommand{\hbar}{\mathchar'26\mkern-9mu h}
\newcommand{\la}{\langle}
\newcommand{\ra}{\rangle}
\newcommand{\e}{\varepsilon}
\newcommand{\w}{\omega}

\begin{document}

\title{Structure of chaotic eigenstates and their entanglement entropy}

\author{Chaitanya Murthy}
\email{cm@physics.ucsb.edu}
\author{Mark Srednicki}
\email{mark@physics.ucsb.edu}
\affiliation{Department of Physics, University of California, Santa Barbara, CA 93106}

\begin{abstract}

We consider a chaotic many-body system (i.e.,~one that satisfies the eigenstate thermalization hypothesis) that is split into two subsystems, with an interaction along their mutual boundary, and study the entanglement properties of an energy eigenstate with nonzero energy density.
When the two subsystems have nearly equal volumes, we find a universal correction to the entanglement entropy that is proportional to the square root of the system's heat capacity (or a sum of capacities, if there are conserved quantities in addition to energy). 
This phenomenon was first noted by Vidmar and Rigol in a specific system; our analysis shows that it is generic, and expresses it in terms of thermodynamic properties of the system. Our conclusions are based on a refined version of a model of a chaotic eigenstate originally due to Deutsch, and analyzed more recently by Lu and Grover.

\end{abstract}

\maketitle

\section{Introduction\label{sec:intro}}

Consider a macroscopic system of volume $V$ partitioned into two spatial subsystems 1 and 2 with volumes $V_1$ and $V_2 = V - V_1$.
We assume, without loss of generality, that $V_1 \leq V_2$.
We also assume that the hamiltonian of the system is a sum of local terms, and so can be  partitioned as
\beq
H = H_1 + H_2 + H_{12},
\label{eq:H}
\eeq
where $H_a$ acts nontrivially only in region $a$ ($a=1,2$) and all terms coupling the two subsystems are contained in $H_{12}$. 
We further assume that the system obeys the eigenstate thermalization hypothesis (ETH) \cite{Deutsch1991,Srednicki1994,Rigol2008} 
for matrix elements of local observables between energy eigenstates corresponding to nonzero energy densities.

To simplify notation, we take all energies to be in units of a fundamental energy scale (e.g., the coefficient of an exchange term in a spin chain) and all lengths, areas, and volumes to be in units of a fundamental length (e.g., the lattice spacing).
We also set $k_B = 1$ throughout.

Let $\ket{E}$ denote an eigenstate of $H$ with energy $E$, with nonzero energy density $E/V$.
For notational convenience, and again without loss of generality, we shift $H_{12}$ by a constant (if necessary) so that
\beq
\la E | H_{12} | E \ra = 0.
\label{eq:H12}
\eeq
We can write $\ket{E}$ in a basis of tensor products of
the eigenstates of $H_1$ and $H_2$,
\beq
\ket{E} = \sum_{i,J} M_{iJ} \ket{i}_1\otimes \ket{J}_2.
\label{eq:E}
\eeq
Deutsch \cite{Deutsch2010} conjectured that the coefficient matrix $M_{iJ}$ can be treated as a random matrix with a narrow bandwidth that keeps the sum of the subsystem energies $E_{1i}+E_{2J}$ close to the total system energy $E$, and using this conjecture showed
that the entanglement entropy of the smaller subsystem equals its thermodynamic entropy.
More recently, Lu and Grover \cite{Lu2019} used this ansatz to calculate the R\'enyi entropies of the subsystem. 
Other related work on entanglement entropy at nonzero energy density in chaotic systems 
includes Refs.~\cite{Santos2012,Deutsch2013,Beugeling2015,Nakagawa2018,Fujita2018};
for a review of basic concepts, see Ref.~\cite{Amico2008}.

In this work, we refine the original conjecture by characterizing the coefficient matrix more completely. 
We further show that at or very near $V_1=V_2$, there is an extra contribution to the entanglement entropy that scales like $\sqrt{V}$.
Specifically, for $V_1=V_2$ exactly, we find that the entanglement entropy is given by
\beq
S_{\text{ent}} = \frac{1}{2} S - \sqrt{\frac{C}{2\pi}} +O(A),
\label{eq:Sent2}
\eeq
where $S$ is the thermodynamic entropy of the system at energy $E$, $C$ is its heat capacity, and $A$ is 
the area of the boundary between the two subsystems. 
When the system is far from a critical point (which we assume for simplicity), both $S$ and $C$ typically scale like the volume $V$ of the system.
We do not compute the coefficient of the $O(A)$ term, since it depends on details of the hamiltonian.
The $\sqrt{C}$ and $O(A)$ terms are distinguished by their scaling with system size (except in $d=2$ spatial dimensions).
They are also distinguished by the fact that the latter depends on a property of the boundary between the two subsystems, while the former depends on a property of the system as a whole.
The extension of Eq.~(\ref{eq:Sent2}) to $V_1 \neq V_2$ is given in Eq.~(\ref{eq:Sent_form}) below; the $\sqrt{C}$ correction remains significant 
for $| V_1 - V_2 | \lesssim \sqrt{V}$.

A contribution to $S_{\text{ent}}$ scaling like $\sqrt{V}$ was found previously by Vidmar and Rigol \cite{Vidmar2017} 
in a study of a one-dimensional system with one conserved quantum number. 
Our explanation for the appearance of such a term is 
essentially the same as theirs, but our formula applies more generally to any system that obeys ETH,
and relates the correction to thermodynamic properties of the system.
Furthermore, we generalize our result to systems with any finite number of conserved quantities in addition to energy.
In such cases, $C$ in Eq.~(\ref{eq:Sent2}) becomes the sum of all entries in a matrix of capacities;
see Sec.~\ref{sec:ee}.

The rest of this paper is organized as follows. 
In the rest of the Introduction, we summarize all of our key results in more precise language, for the simplest case in which only energy is conserved.
Sections~\ref{sec:env}--\ref{sec:ee} elaborate on the derivation of the summarized results.
The generalization to systems with additional conserved quantities is discussed in the second half of Sec.~\ref{sec:ee}, and full details are provided in the \hyperref[app:cons]{Appendix}.
Section~\ref{sec:conclusions} has our concluding discussion.

\vspace{6pt}
\noindent \textbf{$\bullet$ Structure of the coefficient matrix.}
Assuming, in line with Refs.~\cite{Deutsch2010,Lu2019}, that $M_{iJ}$ has the general structure of a random matrix that is sharply banded in total energy, and neglecting any dependence of $M_{iJ}$ on the energy difference $E_{1i} - E_{2J}$, 
we show that it takes the form
\beq
M_{iJ} = e^{-S(E_{1i} + E_{2J})/2} F(E_{1i} + E_{2J} - E)^{1/2} \, C_{iJ} ,
\label{eq:MiJ}
\eeq
where $S(E)$ is the thermodynamic entropy of the full system at energy $E$ (equal to the logarithm of the density of states,
and assumed to be a monotonically increasing function of energy, so that temperature 
is nonnegative), $F(\e)$ is a window function centered on $\e=0$ with a width $\Delta$ equal to the quantum uncertainty in the interaction hamiltonian, 
\beq
\Delta = \sqrt{\la E | H_{12}^2 | E \ra} ,
\label{eq:Delta}
\eeq
and $C_{iJ}$ is a matrix of coefficients which, when averaged over narrow bands of energies of each subsystem near $E_1$ and $E_2$ (but with each band still containing many subsystem energy eigenstates), obeys
\beq
\overline{C_{iJ}}=0, \quad \
\overline{C^*_{iJ}C_{i' J'}}=\delta_{ii'}\delta_{JJ'},
\label{eq:Cstat}
\eeq
where the overbar denotes the dual narrow-band energy averaging.
Furthermore, for a system in two or more spatial dimensions, the window function is a gaussian,
\beq
F(\varepsilon) = \frac{e^{-\varepsilon^2\!/2\Delta^2}}{\sqrt{2\pi} \Delta} .
\label{eq:F}
\eeq
In two or more spatial dimensions, where $H_{12}$ is a sum of local terms along the boundary between the two subsystems, we show that $\Delta \sim \sqrt{A}$, where $A$ is the area of the boundary.
For a one-dimensional system, $\Delta$ is an order-one quantity (in terms of its scaling with system size).

\vspace{6pt}
\noindent \textbf{$\bullet$ Structure of the reduced density matrix.}
The reduced density matrix $\rho_1 \coloneqq \Tr_2 |E\ra\!\la E|$ of subsystem 1 takes the form
\begin{align}
(\rho_1)_{ij} &= e^{-S(E)+S_2(E-E_1)} \bigl[ \delta_{ij} \nonumber \\*[0.2em]
&\hspace{6em} + e^{-S_2(E-E_1)/2} e^{-\w^2\!/8\Delta^2} R_{ij} \bigr] ,
\label{eq:rho12}
\end{align}
where $E_1 \coloneqq (E_{1i} + E_{1j})/2$ and $\w \coloneqq E_{1i} - E_{1j}$, $S_a(E_a)$ is the thermodynamic entropy 
of subsystem $a$ at energy $E_a$ ($a=1,2$), and the $R_{ij}$ are $O(1)$ numbers that vary erratically. We have dropped terms of order $\Delta^2\sim A$ and smaller in the exponents.

The diagonal term is in agreement with Lu and Grover \cite{Lu2019}, and matches the ``subsystem ETH'' ansatz of Dymarsky \textit{et al.}~\cite{Dymarsky2018}.
The off-diagonal term, though exponentially smaller than the diagonal term, alters the spectrum of eigenvalues of $\rho_1$ at energies $E_1>E_1^*$, where $E_1^*$ is the solution to 
\beq
S_1(E_1^*) = S_2(E-E_1^*).
\label{eq:E1*}
\eeq
For $E_1>E_1^*$, the density of states of subsystem 2 is smaller than the density of states of subsystem 1. 
However, the nonzero eigenvalues of $\rho_1$ are the same as those of $\rho_2 \coloneqq \Tr_1 |E\ra\!\la E|$. 
This effect occurs locally in energy.
Hence, in the energy interval $[E_1, E_1 + dE_1]$ for $E_1>E^*_1$, $\rho_1$ has approximately $e^{S_2(E-E_1)} dE_1$ nonzero eigenvalues, and each of these nonzero eigenvalues is approximately equal to $e^{-S(E)} e^{S_1(E_1)}$.

\begin{figure}[t]
    \centering
    \includegraphics[width=0.75\columnwidth]{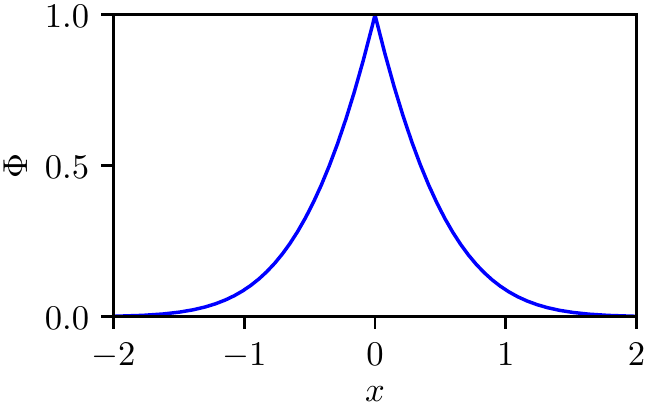}
    \caption{A plot of the function $\Phi(x)$ defined in Eq.~(\ref{eq:Phi}), which parameterizes the correction to the entanglement entropy, Eq.~(\ref{eq:Sent_form}).}
    \label{fig:Phi}
\end{figure}

\vspace{6pt}
\noindent \textbf{$\bullet$ Correction to the entanglement entropy.}
From the discussion above, it follows that 
\beq
\Tr\rho_1^n = \frac{\int dE_1 \, e^{S_{\text{min}}(E_1)} \bigl[e^{-S(E)} e^{S_{\text{max}}(E_1)} \bigr]^n}{e^{-S(E)} \int dE_1  \,e^{S_{\text{min}}(E_1) +S_{\text{max}}(E_1)}},
\label{eq:trrho1n}
\eeq
where
\begin{subequations}
\begin{align}
S_{\text{max}}(E_1) &\coloneqq \max[ S_1(E_1), S_2(E-E_1) ],
\label{eq:Smax} \\
S_{\text{min}}(E_1) &\coloneqq \min[ S_1(E_1), S_2(E-E_1) ] .
\label{eq:Smin}
\end{align}
\end{subequations}
The denominator in Eq.~(\ref{eq:trrho1n}) is the numerator with $n=1$, and itself equals one up to small corrections;
see Sec.~\ref{sec:env}. The entanglement entropy is the $n\to 1$ limit of the $n$th R\'enyi entropy,
\beq
S_{\text{ent}}(E) = \lim_{n\to 1}S_{\text{Ren},n}(E),
\label{eq:Sent}
\eeq
where
\beq
S_{\text{Ren},n}(E) \coloneqq \frac{1}{1-n} \log\Tr\rho_1^n .
\label{eq:SRen}
\eeq
From Eqs.~(\ref{eq:trrho1n})--(\ref{eq:SRen}), we get
\beq
S_{\text{ent}}(E) = \frac{\int \dd{E_1} e^{S_1(E_1) + S_2(E-E_1)} [S(E) - S_{\text{max}}(E_1)]}{\int \dd{E_1} e^{S_1(E_1) + S_2(E-E_1)}} .
\label{eq:Sent_int}
\eeq
After performing the integrals over $E_1$ by Laplace's method, we find
\beq
S_{\text{ent}}(E) = \min(\bar{S}_1, \bar{S}_2) - \sqrt{\frac{2K}{\pi}} \, \Phi\!\left(\frac{\bar{S}_2 - \bar{S}_1}{\sqrt{8 K}}\right) 
+O(A),
\label{eq:Sent_form}
\eeq
where $\bar{S}_1 \coloneqq S_1(\bar{E}_1)$ and $\bar{S}_2 \coloneqq S_2(E-\bar{E}_1)$ are the subsystem entropies at the stationary point $\bar E_1$, given by
\beq
S'_1(\bar E_1)=S'_2(E-\bar E_1),
\label{E1bar}
\eeq
$K \coloneqq C_1 C_2/(C_1+C_2)$ 
is the harmonic mean of the subsystem heat capacities $C_a \coloneqq -\beta^2/\bar S''_a$ 
at constant volume and inverse temperature 
$\beta \coloneqq S'_1(\bar E_1)$,
and we have defined the function
\begin{align}
\Phi(x) 
&\coloneqq \int_{-\infty}^{+\infty}dy\;e^{-y^2} \bigl(|y-x|-|x|\bigr)
\nonumber \\
&\, = \sqrt{\pi} \bigl( x \erf x - \abs{x} \bigr) + e^{-x^2} ,
\label{eq:Phi}
\end{align}
where $\erf x$ is the error function; see Fig.~\ref{fig:Phi}.
Since $\Phi(x)$ decays to zero exponentially from $\Phi(0)=1$, this correction is negligible for $|\bar S_2-\bar S_1|\gg \sqrt{K}$.

For a uniform system with 
$V_1=fV$, $V_2=(1-f)V$, and $f \leq \frac12$,
we have $\bar S_1=fS(E)$, $\bar S_2=(1-f)S(E)$, $C_1=fC$, $C_2=(1-f)C$, and 
$K=f(1-f)C$, where 
$C\coloneqq -\beta^2/S''(E)$ is the heat capacity of the full system. 
The heat capacity $C$ scales like the volume of the system, so 
$|\bar S_2-\bar S_1| \gg \sqrt{K}$ is equivalent to $|\frac12-f| \gg 1/\sqrt{V}$.

For $f=\frac12$ exactly, we recover Eq.~(\ref{eq:Sent2}).

\vspace{6pt}
\noindent \textbf{$\bullet$ Correction to the R\'enyi entropy for $\boldsymbol{n<1}$.}
Evaluating Eq.~(\ref{eq:trrho1n}) by Laplace's method, and then evaluating the leading terms
in Eq.~(\ref{eq:SRen}), we find
\beq
S_{\text{Ren},n}(E) = \frac{\left[S_1({\cal E}_1) + n S_2(E-{\cal E}_1) - n S(E)\right]}{1-n},
\label{eq:SRen2}
\eeq
where
\beq
{\cal E}_1 \coloneqq \min(\bar E_1,E^*_1),
\label{tE1}
\eeq
$E^*_1$ is the solution to Eq.~(\ref{eq:E1*}), and $\bar E_1$ is the solution to
\beq
S'_1(\bar E_1)=n S'_2(E-\bar E_1).
\label{E1barn}
\eeq
For $\bar E_1< E^*_1$, Eq.~(\ref{eq:SRen2}) coincides with the result of Ref.~\cite{Lu2019}.
For $n>1$, the convexity of the entropy function (equivalently, positivity of the temperature and the heat capacity) guarantees that 
$\bar E_1< E^*_1$. 
However, for $n<1$, it is possible to have $E^*_1<\bar E_1$, and then Eq.~(\ref{eq:SRen2}) differs from the result of Ref.~\cite{Lu2019}.
In particular, for a uniform system split exactly in half, $E^*_1<\bar E_1$ for all $n<1$, and then $S_{\text{Ren},n<1}(E) =S(E)/2$, up to subleading corrections.

\section{Envelope function of the coefficient matrix
\label{sec:env}}

We first establish a useful identity. In the limit of $\Delta\to0$, we can ignore the energy of the interaction.
Then we can compute the density of states $e^{S(E)}$ of the total system at energy $E$ by dividing the energy between
the two subsystems, and taking the product of the number of states of each subsystem. This yields
\beq
e^{S(E)} = \int_0^E dE_1\,e^{S_1(E_1)}e^{S_2(E-E_1)}.
\label{eq:eSE}
\eeq
Note that Eq.~(\ref{eq:eSE}) shows that the denominator in Eq.~(\ref{eq:trrho1n}) equals one
in the $\Delta\to0$ limit.

Next we warm up by computing $\la E|E\ra =1$. From Eqs.~(\ref{eq:E}) and (\ref{eq:MiJ}), we have
\beq
\la E|E\ra = \sum_{iJ} e^{-S(E_{1i} + E_{2J})} F(E_{1i} + E_{2J} - E) \, |C_{iJ}|^2. 
\label{eq:EE}
\eeq
The sums over $i$ and $J$ implement the narrow-band averaging of Eq.~(\ref{eq:Cstat}), and can then be replaced by integrals over $E_1$ and $E_2$ with factors of the densities of states, yielding
\begin{align}
\la E|E\ra &= \int_0^\infty dE_1\,e^{S_1(E_1)}\int_0^\infty dE_2\,e^{S_2(E_2)} \nonumber \\*[0.2em]
&\hspace{5em} \times e^{-S(E_{1} + E_{2})} F(E_{1} + E_{2} - E) . 
\label{eq:EE2}
\end{align}
In the limit $\Delta\to 0$, $F(\e)\to\delta(\e)$, the Dirac delta function.
In this limit we have
\begin{align}
\la E|E\ra &= e^{-S(E)}\int_0^\infty dE_1\,e^{S_1(E_1)}e^{S_2(E-E_1)} 
\nonumber \\*
&=1,
\label{eq:EE3}
\end{align}
where the final result follows from Eq.~(\ref{eq:eSE}). 

For finite $\Delta$, we take $F(\e)$ to have the gaussian form of Eq.~(\ref{eq:F}), 
although we only need that $F(\e)$ be sharply peaked at $\e=0$ with width $\Delta$. 
We then evaluate the integrals in Eq.~(\ref{eq:EE2}) by Laplace's method. 
The conditions for a stationary point of the exponent are
\begin{subequations}
\begin{align}
S_1'(E_1) &= S'(E_1 + E_2) + (E_1 + E_2 - E)/\Delta^2 ,
\label{eq:stat1} \\*
S_2'(E_2) &= S'(E_1 + E_2) + (E_1 + E_2 - E)/\Delta^2 .
\label{eq:stat2}
\end{align}
\end{subequations}
For small $\Delta$, the solution is $E_1=\bar E_1$, $E_2=\bar E_2$, where
\begin{subequations}
\label{eq:statE}
\begin{align}
\bar E_1+\bar E_2 &= E,
\label{eq:E1E2E} \\*
S_1'(\bar E_1)=S_2'(\bar E_2) &= S'(E) \eqqcolon \beta ,
\label{eq:Sprimes}
\end{align}
\end{subequations}
where $\beta$ is again the inverse temperature of the system as a whole and of each subsystem.
Next we Taylor-expand the entropies about the stationary point,
\begin{align}
S(E_1 + E_2) &= S(E) + \beta (E_1 + E_2 - E) \nonumber \\*
&\hspace{1.2em} - \tfrac{1}{2} (\beta^2/C) (E_1 + E_2 - E)^2 + \cdots , 
\label{eq:SE1E2} \\*[0.2em]
S_1(E_1) &= \bar S_1 + \beta (E_1 - \bar{E}_1)  \nonumber \\*
&\hspace{1.2em} - \tfrac{1}{2} (\beta^2/C_1) (E_1 - \bar{E}_1)^2 + \cdots , 
\label{eq:S1E1} \\*[0.2em]
S_2(E_2) &= \bar S_2 + \beta (E_2 - \bar{E}_2) \nonumber \\*
&\hspace{1.2em} - \tfrac{1}{2} (\beta^2/C_2) (E_2 - \bar{E}_2)^2 + \cdots .
\label{eq:S2E2}
\end{align}
To leading order in the system volume, Eq.~(\ref{eq:eSE}) implies
\beq
\bar S_1+\bar S_2 = S(E).
\label{eq:Ssum}
\eeq
Thus the constant and linear terms all cancel in the combination $S_1(E_1)+S_2(E_2)-S(E_1+E_2)$
that appears in Eq.~(\ref{eq:EE2}). This cancellation is why it was necessary to have $S(E_1+E_2)$
in the exponent in Eq.~(\ref{eq:MiJ}) rather than $S(E)$.
The remaining quadratic terms yield gaussian integrals that give
an $O(1)$ result for the integral in Eq.~(\ref{eq:EE2}). Adjusting the $O(1)$ terms in the entropies is then
necessary to yield the final result of $\la E|E\ra=1$.

The heat capacities in Eqs.~(\ref{eq:SE1E2})--(\ref{eq:S2E2}) are proportional to the volumes of the corresponding macroscopic regions, whereas $\Delta^2 \propto A$ (the area of the 1--2 boundary), as will be demonstrated below. 
Consequently $1/\Delta^2 \gg \beta^2/C, \beta^2/C_1, \beta^2/C_2$,
and hence the distribution of $E_1+E_2$ is controlled by $F(\e)$. Then 
we have the following generalization of Eq.~(\ref{eq:EE2}),
\beq
\la E|(H_1+H_2-E)^n|E\ra \approx \int_{-\infty}^{+\infty}d\e\, F(\e)\e^n.
\label{eq:HHEn}
\eeq
We emphasize that the derivation of Eq.~(\ref{eq:HHEn}) from Eqs.~(\ref{eq:MiJ}) and (\ref{eq:Cstat}) does not rely on the precise form of $F(\e)$; it relies only on $F(\e)$ being sharply peaked at $\e = 0$ with width $\Delta$ that satisfies $\beta^2 \Delta^2 \ll C_1, C_2$.

For $n=1,2$ in Eq.~(\ref{eq:HHEn}), we can replace $H_1+H_2-E$ with $H_1+H_2-H$, since $H$
will always appear next to either the ket or bra form of its eigenstate.
Then using $H_1+H_2-H=-H_{12}$, we find for $n=1$ that 
\beq
\int_{-\infty}^{+\infty}d\e\,F(\e)\e = -\la E|H_{12}|E\ra = 0,
\label{eq:ave}
\eeq
where the second equality follows from our shift of $H_{12}$.
For $n=2$, we get
\beq
\int_{-\infty}^{+\infty}d\e\,F(\e)\e^2 = \la E|H_{12}^2|E\ra.
\label{eq:D2}
\eeq
The left-hand side equals $\Delta^2$ by definition, and so Eq.~(\ref{eq:D2})
verifies Eq.~(\ref{eq:Delta}).

In two or more spatial dimensions, our assumption on the locality of $H$
implies that $H_{12}$ is a sum of local terms on the boundary $B$ between regions 1 and 2,
\beq
H_{12} = \sum_{x\in B} h_x .
\label{eq:H12sum}
\eeq
We then have 
\beq
\la E|H_{12}^2|E\ra = \sum_{x,y \in B } \la E|h_x h_y|E\ra.
\label{eq:H122}
\eeq
Assuming that ETH holds for the bilocal operator $h_x h_y$, 
the eigenstate expectation value can be replaced by a thermal expectation value at
inverse temperature $\beta$. We further assume that this thermal correlation function
decays rapidly for $|x-y|\gg\xi$ to the disconnected form $\la h_x\ra\la h_y\ra$, 
where $\xi$ is an appropriate correlation length
\footnote{Note that the eigenstate expectation value 
\unexpanded{$\langle E | h_x h_y | E \rangle$} 
must in general differ from the thermal expectation value 
\unexpanded{$\langle h_x h_y \rangle$} 
by $O(1/V)$, even when $|x-y| \gg \xi$; this is needed to recover 
\unexpanded{$\langle E | (H - E)^2 | E \rangle = 0$}. 
However, this difference only contributes an $O(A/V)$ correction to Eq.~(\ref{eq:H123}), and hence can be neglected.}.
Summing the disconnected form
over $x$ and/or $y$ yields zero, by Eq.~(\ref{eq:H12}). Hence the double sum in 
Eq.~(\ref{eq:H122}) effectively becomes a single sum over the boundary, yielding
\beq
\Delta^2 = \la E|H_{12}^2|E\ra \sim A \, \xi^{d-1} \la h_x^2\ra,
\label{eq:H123}
\eeq
where $h_x$ is any one term in $H_{12}$, and the angle brackets denote either the eigenstate or thermal average, which are equal by ETH. 
Equation~(\ref{eq:H123}) shows that $\Delta^2\sim A$, the boundary area.

We can now generalize this argument to higher powers of $H_{12}$, again assuming 
rapid decay of $\la h_x h_y \cdots\ra$ whenever an index or group of indices is separated
by more than $\xi$ from the others. The multiple sum over $x,y,\ldots$ will then yield
approximately zero for odd powers, and be dominated by the factorization into correlated pairs
for even powers
\footnote{
More precisely, the third and higher cumulants of $H_{12}/\Delta$ in the state \unexpanded{$| E \rangle$} are suppressed relative to the variance, \unexpanded{$\langle E | (H_{12}/\Delta)^2 | E \rangle \equiv 1$}, by powers of $1/\sqrt{A}$.
}. 
This then yields, in accord with the usual combinatorics of Wick's theorem,
\beq
\la E|H_{12}^{2n}|E\ra \approx (2n-1)!! \,\Delta^{2n},
\label{eq:H12n}
\eeq
characteristic of a gaussian distribution. 

Returning to Eq.~(\ref{eq:HHEn}), and using
\beq
H_1+H_2-E=H-E-H_{12},
\label{eq:H1H2}
\eeq
we have
\begin{align}
\int_{-\infty}^{+\infty} d\e\, F(\e)\e^{2n}
&\approx \ev{(H - E - H_{12})^{2n}} 
\nonumber \\*
&\approx \ev{H_{12}^{2n}} 
- \ev{H_{12} (H-E) H_{12}^{2n-2}} 
+ \cdots ,
\label{eq:Fmom}
\end{align}
The first term is given by Eq.~(\ref{eq:H12n}), and we would like to show that the remaining terms
can be neglected. From Eqs.~(\ref{eq:H123}) and (\ref{eq:H12n}), we see that we effectively have
$H_{12}\sim \sqrt{A}$, so the terms in Eq.~(\ref{eq:Fmom}) with factors of $H-E$ will be suppressed
unless $H-E\sim \sqrt{A}$ as well. In each of these terms, $H$ acts on a state of the form
$H_{12}^k|E\ra$. We have $H_{12}=\sum_x h_x$, and each $h_x$ is an $O(1)$ operator
that can change the energy only by an $O(1)$ amount. Hence, acting with $k$ such operators can change the 
energy by at most an $O(k)$ amount, which is $O(1)$ in terms of its scaling with $A$. 
Summing over $x$ can increase the coefficient of the normalized state, but does not increase the maximum change in energy.
Hence $H-E \sim O(1)$, and so the terms with one or more factors of $H-E$ in Eq.~(\ref{eq:Fmom}) can be neglected.
We conclude that, up to corrections suppressed by powers of $\xi^{d-1}/A$ or $A/V$,
\beq
\int_{-\infty}^{+\infty} d\e\, F(\e)\e^{2n} =  (2n-1)!! \,\Delta^{2n},
\label{eq:Fmom2}
\eeq
and therefore that $F(\e)$ is a gaussian with width $\Delta$, Eq.~(\ref{eq:F}).

In one spatial dimension, $H_{12}$ is a single term, rather than a sum of $O(A)$ terms.
Hence the combinatoric analysis that led to Eq.~(\ref{eq:H12n}) does not apply,
and so we cannot conclude that the shape of $F(\e)$ is gaussian.
However, Eqs.~(\ref{eq:ave}) and (\ref{eq:D2}) are still valid, and so $F(\e)$
is still sharply peaked at $\e=0$ with a width $\Delta$ that is given by Eq.~(\ref{eq:Delta}).
All of our results for the corrections to the entanglement entropy, including Eqs.~(\ref{eq:Sent2}) and (\ref{eq:Sent_form}),
and their generalizations to multiple conserved quantities via Eqs.~(\ref{eq:K1})--(\ref{eq:C1}) below,
only depend on this sharply-peaked nature of $F(\e)$, and not on the details of its shape,
and so hold for all dimensions, including $d=1$.

\section{Reduced density matrix}
\label{sec:rdm}

From Eq.~(\ref{eq:E}), the reduced density matrix of subsystem 1 is 
\beq
(\rho_1)_{ij} = \sum_K  M_{iK}M^*_{jK}.
\label{eq:rho1ij}
\eeq
Using Eqs.~(\ref{eq:MiJ}), (\ref{eq:F}), and (\ref{eq:SE1E2}), assuming $\Delta^2\ll C/\beta^2$, and neglecting prefactors, we have
\begin{align}
(\rho_1)_{ij} &= e^{-S(E) - \omega^2/8\Delta^2} \sum_K \Bigl[ e^{-\beta (E_{2K} + E_1 - E)} 
\nonumber \\*
&\hspace{5em} \times e^{-(E_{2K} + E_1 - E)^2/2\Delta^2}  C_{iK} C_{jK}^* \Bigr],
\label{eq:rho1ij2}
\end{align}
where $E_1 \coloneqq (E_{1i} + E_{1j})/2$ and $\w \coloneqq E_{1i} - E_{1j}$. 
Taking the statistical average and using Eq.~(\ref{eq:Cstat}), only the diagonal term survives, hence $\w=0$, and we get
\begin{align}
\overline{(\rho_1)_{ij}} = 
e^{-S(E)} \sum_K \Bigl[&
e^{-\beta (E_{2K} + E_1 - E)} 
\nonumber \\*[-0.6em]
&\times e^{-(E_{2K} + E_1 - E)^2/2\Delta^2} \Bigr] \delta_{ij}.
\label{eq:barrho1}
\end{align}
We again replace the sum over $K$ with an integral over $E_2$ weighted by the density of states
of subsystem 2, which yields
\begin{align}
\overline{(\rho_1)_{ij}} = 
e^{-S(E)} \int_0^\infty dE_2 \, &e^{S_2(E_2)}
e^{-\beta (E_{2} + E_1 - E)} 
\nonumber \\*
&\times e^{-(E_{2} + E_1 - E)^2/2\Delta^2} \delta_{ij}.
\label{eq:barrho2}
\end{align}
The last exponential factor, arising from the window function $F(\e)$, forces $E_2$ to be close to $E-E_1$.
Expanding $S_2(E_2)$ about this point, we have
\beq
S_2(E_2) = S_2(E-E_1) +\beta_{21}(E_2+E_1-E) + \cdots,
\label{eq:s2e2}
\eeq
where $\beta_{21}\coloneqq S'_2(E-E_1)$ is the inverse temperature of subsystem 2 when its energy is $E-E_1$.
Performing the integral over $E_2$ in Eq.~(\ref{eq:barrho2}) then yields
\beq
\overline{(\rho_1)_{ij}} = 
e^{-S(E) +S_2(E-E_1)+\Delta^2(\beta-\beta_{21})^2/2} \delta_{ij}.
\label{eq:barrho3}
\eeq
Since $\Delta^2\sim A$, the last term in the exponent is smaller than the first two,
which scale like volume. Additionally, we expect other terms of $O(A)$ to arise from
finer structure in the $C_{iJ}$ coefficients that we have neglected.
These are necessary to produce the usual ``area law'' for the entanglement entropy
of the ground state (for a review, see Ref.~\cite{Eisert2010}), 
and we expect such correlations to persist at nonzero energy density.

To estimate the size of the fluctuating off-diagonal elements of $\rho_1$, we compute
the statistical average of the absolute square of $(\rho_1)_{ij}$, $i\ne j$.
We neglect any statistical correlations in the $C_{iK}$ coefficients, and assume that 
\beq
\overline{C_{iK}C^*_{jK}C^*_{iL}C_{jL}} =\delta_{KL}.
\label{eq:CCCC}
\eeq
Then we have
\begin{align}
\overline{|(\rho_1)_{ij}|^2} = 
e^{-2S(E) - \omega^2/4\Delta^2} &\sum_K \Bigl[
e^{-2\beta (E_{2K} + E_1 - E)} 
\nonumber \\*
&\, \times e^{-(E_{2K} + E_1 - E)^2/\Delta^2} \Bigr].
\label{eq:barrhosq}
\end{align}
Following the same steps that led to Eq.~(\ref{eq:barrho3}), we get
\beq
\overline{|(\rho_1)_{ij}|^2} = 
e^{-2S(E) +S_2(E-E_1) - \omega^2\!/4\Delta^2 + \Delta^2(\beta-\beta_{21}/2)^2} .
\label{eq:barrhosq2}
\eeq
As in the case of the diagonal components, we expect additional terms of $O(A)$ to arise from
neglected correlations in the $C_{iK}$ coefficients.

In the limit that we neglect $O(A)$ corrections, 
Eqs.~(\ref{eq:barrho3}) and (\ref{eq:barrhosq2}) together yield Eq.~(\ref{eq:rho12}).

\section{Corrections to the entanglement and R\'enyi entropies}
\label{sec:ee}

In the limit that we neglect all subleading corrections, we evaluate 
the numerator of Eq.~(\ref{eq:trrho1n}) by Laplace's method, which simply yields the maximum value
of the integrand. This gives Eq.~(\ref{eq:SRen2}) for the $n$th R\'enyi entropy.

We consider subleading corrections only in the case of the entanglement entropy, $n=1$.
In this case we must evaluate Eq.~(\ref{eq:Sent_int}).
We have
\beq
S_{\text{max}} = {\textstyle\frac12} (S_1+S_2) + {\textstyle\frac12} |S_1-S_2|.
\label{eq:Smax2}
\eeq
Using Eqs~(\ref{eq:E1E2E}) and (\ref{eq:S1E1})--(\ref{eq:Ssum}),
and changing the integration variable from $E_1$ to 
\beq
u\coloneqq \beta(E_1-\bar E_1),
\label{eq:u}
\eeq
we get
\begin{align}
S_1+S_2 &= S - u^2/2K,
\label{eq:S1pS2} \\
S_1-S_2 &=  \bar S_1-\bar S_2 + 2u +O(u^2/K),
\label{eq:S1mS2}
\end{align}
where again $K:=C_1 C_2/(C_1+C_2)$.
The factor of $e^{-u^2/2K}$ in the integrand is peaked well away from the lower limit of integration, 
which can therefore be extended to $-\infty$.
When performing the gaussian integral, values of $u^2$ larger than $K$ are exponentially suppressed; 
thus the $O(u^2/K)$ term in Eq.~(\ref{eq:S1mS2}) gives only an $O(1)$ contribution, and can be neglected.
Putting all of this together, Eq.~(\ref{eq:Sent_int}) becomes
\beq
S_{\text{ent}} =\tfrac{1}{2} S 
- \frac{\int_{-\infty}^{+\infty} \dd{u} e^{-u^2/2K} \abs{\frac{1}{2}(\bar{S}_2 - \bar{S}_1) - u}}{\int_{-\infty}^{+\infty} \dd{u} e^{-u^2/2K}} .
\label{eq:SentK}
\eeq
Making a final rescaling of $u\to\sqrt{2K}y$, we get Eq.~(\ref{eq:Sent_form}).

Note that, if we are interested in infinite temperature ($\beta = 0$), then we should also take $C_j \to 0$ so that $\beta^2/C_j$ remains finite and nonzero. In this limit, $C\to 0$, and so the correction in Eq.~(\ref{eq:Sent2}) vanishes.

We also note that the precise distribution of eigenvalues of $\rho_1$ near $E_1$ makes at most an $O(1)$ correction to the entanglement entropy.
For example, the Mar{\v{c}}enko-Pastur law \cite{Marcenko1967} gives the Page correction 
$-e^{S_{\text{min}}}/2e^{S_{\text{max}}}$ 
to the entanglement entropy of a random state \cite{Page1993,Bianchi2019}.
This can be neglected.

We can also generalize to the case of a system with another conserved quantity, such as particle number.
In the most general case, there are $m$ conserved quantities $Q^a$ ($a=1,\ldots,m$), including energy, which we take to be $Q^1$. 
A quantum state is then labeled by the values of all $m$ quantities. We can then repeat our entire analysis (see the \hyperref[app:cons]{Appendix} for details). 
The thermodynamic entropy of the full system as a function of the $Q^a$'s (near the values that label the state) takes the form 
\begin{align}
S(Q+\delta Q) &= S(Q) + \lambda^a \delta Q^a \nonumber \\*
&\quad - \tfrac{1}{2}(\mathbf{C}^{-1})_{ab} \, \lambda^a \delta Q^a \lambda^b \delta Q^b + \cdots,
\label{eq:SQ}
\end{align}
with $\lambda^1 \equiv \beta$. This generalizes Eq.~(\ref{eq:SE1E2}); similar generalizations apply to the subsystem entropies.
We then ultimately arrive at Eq.~(\ref{eq:SentK}) with $u \coloneqq \lambda^a \delta Q^a$ and
\begin{align}
K &\coloneqq \sum_{a,b=1}^m \left[ (\mathbf{C}_1^{-1} + \mathbf{C}_2^{-1})^{-1} \right]_{ab}
\label{eq:K1} \\
&\, = f(1-f) \sum_{a,b=1}^m \mathbf{C}_{ab} ,
\label{eq:K2}
\end{align}
where $\mathbf{C}_1$ and $\mathbf{C}_2$ are the capacity matrices for the two subsystems,
and the second equality holds for a uniform system
with capacity matrix $\mathbf{C}$ and with $f= V_1/V$.
Equation~(\ref{eq:Sent_form}) then holds with $K$ given by Eq.~(\ref{eq:K1}), and
Eq.~(\ref{eq:Sent2}) holds with 
\beq
C = \sum_{a,b=1}^m \mathbf{C}_{ab} .
\label{eq:C1}
\eeq

We can now reproduce the results of Ref.~\cite{Vidmar2017} for a system with a conserved particle number.
There the system was studied near infinite temperature, so that the thermodynamic entropy was taken to be
effectively independent of system energy. Hence the problem reduces to the case of a single conserved quantity,
the filling fraction $n$. Then the thermodynamic entropy of the system takes the form
\beq
S(n) = -L \, [n\ln n + (1-n)\ln(1-n)],
\label{eq:Sn}
\eeq
where $L$ is the linear volume of the one-dimensional system; this is Eq.~(13) in Ref.~\cite{Vidmar2017}.
In the notation of our Eq.~(\ref{eq:SQ}), with a single $Q$ that we identify as $n$, we have
\begin{align}
\lambda &= S'(n),
\label{eq:lam} \\
\lambda^2 C^{-1} &= -S''(n),
\label{eq:C}
\end{align}
which yields
\beq
C = L \, n(1-n)\left[\ln\left(\frac{1-n}n\right)\right]^2.
\label{eq:C2}
\eeq
When used in Eq.~(\ref{eq:Sent2}), this reproduces Eq.~(17) of Ref.~\cite{Vidmar2017} (with $L_A=L/2$).

\section{Conclusions}
\label{sec:conclusions}

We have reconsidered the ansatz of Refs.~\cite{Deutsch2010,Lu2019} for an energy eigenstate of a chaotic many-body system that, by assumption, obeys the eigenstate thermalization hypothesis for local observables. 
This ansatz expresses the energy eigenstate of the full system in the basis of energy eigenstates of two subsystems, each contiguous in space, that interact along their mutual boundary, and is specified by Eqs.~(\ref{eq:E}), (\ref{eq:MiJ}), and (\ref{eq:Cstat}).

One of the results of this paper is that the width $\Delta$ of the energy window function $F(\e)$ is given
by Eq.~(\ref{eq:Delta}) in terms of the subsystem interaction hamiltonian, 
and that (in two or more spatial dimensions) $F(\e)$ has the gaussian form of Eq.~(\ref{eq:F}).

We further showed that the ansatz for the energy eigenstate leads to a reduced density matrix that takes the 
form of Eq.~(\ref{eq:rho12}). 
The off-diagonal elements, though exponentially small, are relevant to the calculation of R\'enyi entropies 
when the fraction of the energy in the smaller subsystem is large enough to give it a larger entropy than the larger subsystem;
this modifies the results of Ref.~\cite{Lu2019} for $n<1$.

In the case of equal or nearly equal volume for the two subsystems,
there is a correction to the entanglement entropy (corresponding to R\'enyi index $n=1$) 
that scales like the square-root of the system volume. In the case of equal subsystem volumes, this correction,
displayed in Eq.~(\ref{eq:Sent2}), is $\Delta S_\text{ent}=-\sqrt{C/2\pi}$, where $C$ is the heat capacity of the whole system. 
Such a correction was previously found in a specific system by Vidmar and Rigol \cite{Vidmar2017};
our analysis is more general and shows that the effect is generic. 

We also extended our results to the case of multiple conserved quantities. The correction to the entanglement entropy at equal subsystem volumes is the same, but with $C$ now given by a sum of the elements of a matrix of capacities.

We believe that our work further illuminates the role of the entanglement and R\'enyi entropies of a subsystem as
quantities worthy of study that encode key features of the physical properties of the system as a whole.

\begin{acknowledgments}
We thank Eugenio Bianchi, Tarun Grover, Tsung-Cheng Lu, Marcos Rigol, and Lev Vidmar for helpful discussions. This work was supported in part by the Microsoft Corporation Station Q (C.M.).
\end{acknowledgments}

\appendix

\setcounter{equation}{0}
\renewcommand{\theequation}{A\arabic{equation}}

\section*{Appendix: Multiple conserved quantities}
\label{app:cons}

In the most general case, there are $m$ conserved quantities $Q^a$ ($a=1,\ldots,m$), including energy, which we take to be $Q^1$.
We assume that each $Q^a$ is a sum of local terms, and so can be partitioned as in Eq.~(\ref{eq:H}),
\beq
Q^a = Q^a_1 + Q^a_2 + Q^a_{12} ,
\label{eq:Q}
\eeq
with
\begin{subequations}
\label{eq:Qcomm}
\begin{align}
\comm{Q^a}{Q^b} &= 0 , \\*
\comm{Q^a_1}{Q^b_1} &= 0 , \\*[0.1em]
\comm{Q^a_2}{Q^b_2} &= 0 .
\end{align}
\end{subequations}
Let $\ket{q}$ denote a simultaneous eigenstate of the $Q^a$, with eigenvalues $q^a$.
Without loss of generality, we shift each $Q^a_{12}$ so that
\beq
\la q | Q^a_{12} | q \ra = 0.
\eeq

We can write $\ket{q}$ in a basis of tensor products of the eigenstates of $Q^a_1$ and $Q^a_2$,
\beq
\ket{q} = \sum_{i,J} M_{iJ} \ket{i}_1\otimes \ket{J}_2,
\eeq
where $Q^a_1 \ket{i}_1 = q^a_{1i} \ket{i}_1$ and $Q^a_2 \ket{J}_2 = q^a_{2J} \ket{J}_2$.
The thermodynamic entropy of the full system is now a function $S(q) \equiv S(q^1,\dots,q^m)$ of all the $q^a$'s.
Its Taylor expansion, about the values that label the state, takes the form
\begin{align}
S(q+\delta q) &= S(q) + \lambda^a \delta q^a
\nonumber \\
&\quad - \tfrac{1}{2} (\mathbf{C}^{-1})_{ab} \, \lambda^a \delta q^a \lambda^b \delta q^b + \cdots,
\end{align}
with $\lambda^1 \equiv \beta$. This generalizes Eq.~(\ref{eq:SE1E2}).
We assume that $\lambda^a > 0$ and that the capacity matrix $\mathbf{C}$ is positive definite; 
this generalizes positivity of temperature and heat capacity.
Similar generalizations apply to the subsystem entropies.

We can then repeat our entire analysis.
The coefficient matrix $M_{iJ}$ takes the form
\beq
M_{iJ} = e^{-S(q_{1i} + q_{2J})/2} F(q_{1i} + q_{2J} - q)^{1/2} \, C_{iJ} .
\eeq
Here $F(z)$ is a window function centered on $z_a=0$ with second moments given by
\beq
\int \dd[m]{z} F(z) \, z_a  z_b = \mathbf{D}_{ab} \coloneqq \la q | Q_{12}^a Q_{12}^b | q \ra .
\label{eq:Fq}
\eeq
Equations~(\ref{eq:Q}) and (\ref{eq:Qcomm}) together imply
\beq
\la q | \comm{Q^a_{12}}{Q^b_{12}} | q \ra = 0 ,
\label{eq:Q12comm}
\eeq
so $\mathbf{D}_{ab}$ is symmetric, as it needs to be for the equality in Eq.~(\ref{eq:Fq}) to make sense.
The $C_{iJ}$ coefficients obey Eq.~(\ref{eq:Cstat}), with the averaging now over narrow bands of all components of 
$q_1$ and $q_2$.
For a system in two or more spatial dimensions, the window function is a multivariate gaussian,
\beq
F(z) = \frac{e^{- z \cdot \mathbf{D}^{-1} z/2}}{(2\pi)^{m/2} \sqrt{\det \mathbf{D}}} .
\eeq
The matrix elements of $\mathbf{D}$ scale like $\mathbf{D}_{ab} \sim A$, where $A$ is the area of the boundary between regions 1 and 2.
When $m = 1$, $\mathbf{D}$ reduces to $\Delta^2$.

The reduced density matrix $\rho_1 \coloneqq \Tr_2 |q\ra\!\la q|$ of subsystem 1 takes the form
\begin{align}
(\rho_1)_{ij} &= e^{-S(q)+S_2(q-q_1)} \bigl[ \delta_{ij} \nonumber \\*[0.2em]
&\hspace{5em} + e^{-S_2(q-q_1)/2} e^{- w \cdot \mathbf{D}^{-1} w/8} R_{ij} \bigr] ,
\label{eq:rho12_q}
\end{align}
where $q^a_1 \coloneqq (q^a_{1i} + q^a_{1j})/2$ and $w^a \coloneqq q^a_{1i} - q^a_{1j}$, the $R_{ij}$ are $O(1)$ numbers that vary erratically, and we have dropped terms of order $\mathbf{D}_{ab} \sim A$ in the exponents.

We adopt the notation
\begin{subequations}
\begin{align}
q_1 &\prec q_1^* \quad \text{if} \quad
S_1(q_1) < S_2(q-q_1) , \\
q_1 &\succ q_1^* \quad \text{if} \quad
S_1(q_1) > S_2(q-q_1) , \\
q_1 &\sim q_1^* \quad \text{if} \quad
S_1(q_1) = S_2(q-q_1) .
\end{align}
\end{subequations}
In other words, $\succ$ is the order on $q_1$ induced by the function $S_1(q_1) - S_2(q-q_1)$, and $q_1^*$ is some point at which this function vanishes.

The off-diagonal term in Eq.~(\ref{eq:rho12_q}), though exponentially smaller than the diagonal term, is relevant for $q_1 \succ q_1^*$.
In the small box $[q^1_1, q^1_1 + dq^1_1] \times \cdots \times [q^m_1, q^m_1 + dq^m_1]$ for $q_1 \succ q^*_1$, $\rho_1$ has approximately $e^{S_2(q-q_1)} dq^1_1 \cdots dq^m_1$ nonzero eigenvalues, each one approximately equal to $e^{-S(E)} e^{S_1(q_1)}$.

Equation~(\ref{eq:SRen2}) for the R\'enyi entropy generalizes to
\beq
S_{\text{Ren},n}(q) = \frac{\left[S_1(\mathcal{Q}_1) + n S_2(q-\mathcal{Q}_1) - n S(q)\right]}{1-n} ,
\eeq
where $\mathcal{Q}_1$ is the point at which $S_1(q_1) + n S_2(q-q_1)$ attains its maximal value in the region $q_1 \precsim q_1^*$.
For $n>1$, the convexity of the entropy function guarantees that $\mathcal{Q}_1\prec q^*_1$. 
In this case, $\mathcal{Q}_1 = \bar{q}_1$, the solution to
\beq
\nabla S_1(\bar{q}_1) = n \nabla S_2(q-\bar{q}_1) .
\eeq
However, for $n<1$, it is possible to have $\mathcal{Q}_1 \sim q_1^* \prec \bar{q}_1$.
In particular, for a uniform system split exactly in half, $q^*_1\prec \bar{q}_1$ for all $n<1$, and then $S_{\text{Ren},n<1}(q) =S(q)/2$, up to subleading corrections.

For the entanglement entropy, we repeat the steps in Sec.~\ref{sec:ee} and arrive at the generalization of Eq.~(\ref{eq:SentK}),
\beq
S_{\text{ent}} =\tfrac{1}{2} S
- \frac{\int \dd[m]{v} e^{- v \cdot \mathbf{K}^{-1} v/2} \abs{\frac{1}{2}(\bar{S}_2 - \bar{S}_1) - r \cdot v}}{\int \dd[m]{v} e^{-v \cdot  \mathbf{K}^{-1} v /2}} ,
\label{eq:SentK_mat}
\eeq
where $\mathbf{K}^{-1} \coloneqq \mathbf{C}_1^{-1} + \mathbf{C}_2^{-1}$, $\mathbf{C}_1$ and $\mathbf{C}_2$ are the capacity matrices for the two subsystems, $r \coloneqq (1,1,\dots,1)$, $v^a \coloneqq \lambda^a (q^a_1 - \bar{q}^a_1)$, $\bar{S}_1 \coloneqq S_1(\bar{q}_1)$, $\bar{S}_2 \coloneqq S_2(q-\bar{q}_1)$, and $\bar{q}_1$ is the solution to
\beq
\nabla S_1(\bar{q}_1) = \nabla S_2(q-\bar{q}_1) .
\eeq
\linebreak
The integral in Eq.~(\ref{eq:SentK_mat}) is of the form
\begin{align}
I &= \int \dd[m]{v} \, e^{- v \cdot \mathbf{K}^{-1} v/2} f(r \cdot v) \nonumber \\*
&= \int \dd{u} \int \dd[m]{v} \, e^{- v \cdot \mathbf{K}^{-1} v/2} \delta(u - r \cdot v) \, f(u) \nonumber \\*
&= \int \dd{u} \int \frac{\dd{k}}{2\pi} \int \dd[m]{v} \, e^{- v \cdot \mathbf{K}^{-1} v/2 + ik(u - r \cdot v)} f(u) .
\end{align}
Performing all the gaussian integrals,
\beq
I \propto \int \dd{u} e^{-u^2 / (2 \, r \cdot \mathbf{K} r)} f(u) .
\eeq
Thus, Eq.~(\ref{eq:SentK_mat}) reduces to Eq.~(\ref{eq:SentK}) with
\beq
K = r \cdot \mathbf{K} \, r 
= \sum_{a,b=1}^m \mathbf{K}_{ab} ,
\label{eq:Keff}
\eeq
which is equivalent to Eq.~(\ref{eq:K1}).

\end{document}